\documentclass[%
reprint,
 amsmath,amssymb,mathtools
 aps,
]{revtex4-2}

\usepackage{graphicx}
\usepackage{xcolor}
\usepackage{dcolumn}
\usepackage{bm}
\usepackage[colorlinks=true, linkcolor=blue, citecolor=blue, urlcolor=blue]{hyperref}
\usepackage[mathlines]{lineno}

\newcommand{\abs}[1]{\ensuremath{\left\vert#1\right\vert}}	

\begin{document}

\preprint{APS/123-QED}

\title{Amoeboid cell migration and shape dynamics driven by actin polymerization}

\author{Winfried Schmidt}
\affiliation{%
	Univ.~Grenoble Alpes, CNRS, LIPhy, 38000 Grenoble, France
}%
\affiliation{%
	Centre national d’études spatiales (CNES), Paris, France 
}%
\author{Chaouqi Misbah}
\affiliation{%
	Univ.~Grenoble Alpes, CNRS, LIPhy, 38000 Grenoble, France
}%
\author{Alexander Farutin}
\affiliation{%
	Univ.~Grenoble Alpes, CNRS, LIPhy, 38000 Grenoble, France
}%

\date{\today}

\begin{abstract}
	Cell migration is fundamental to development, tissue organization, immune response, and disease progression. Amoeboid motility is distinguished by rapid motion and strongly fluctuating cell shapes, reflecting the intrinsically nonlinear nature of active living matter far from equilibrium. Here we introduce a minimal active-shell model of an amoeboid cell that couples actin polymerization, cortical flows, and membrane deformation through nonlocal mechanical interactions. The model gives rise to a rich spectrum of emergent behaviors. A symmetric non-motile state can spontaneously break symmetry and transition toward persistent directed migration driven solely by polymerization-induced retrograde flow, even in the absence of shape deformation. Increasing activity further triggers a cascade of dynamical states, including circular trajectories, oscillatory zigzag motion, and irregular chaotic-like migration with fluctuating protrusions and multi-lobed morphologies. Although these migratory modes are observed experimentally in distinct cellular contexts, our results show that they can emerge from the same underlying physical mechanism, providing a unified framework for amoeboid dynamics. Notably, contractile stresses induced by molecular motors are not required to generate spontaneous motility, polarity, or complex migration patterns. Our findings highlight how collective active processes at the cellular scale can self-organize into complex dynamical states, revealing generic principles of nonlinear behavior in living systems.
\end{abstract}

\maketitle

\section{Introduction}
\label{sec:intro}
Cell migration is crucial for a variety of biological processes, such as the immune system, wound healing, embryogenesis, and cancer metastasis.
Two main migration modes have been identified: The mesenchymal mode which is slow and relies on specific adhesion, observed for instance for fibroblasts, and the amoeboid mode which is fast and is characterized by weak adhesion and highly dynamic cell shapes, observed for leukocytes and cancer cells.
Some cell types can also switch between the two migration modes, depending on their environment.
Amoeboid motility is driven by a tangential retrograde flow of the actin cortex which generates the forces necessary for locomotion. It is driven by polymerization of cortical actin filaments primarily at the cell front in combination with the activity of myosin, molecular motors that accumulate in the cell rear where they contract the cortex.
The actin retrograde flow has been shown to drive amoeboid cell migration in various environments, including adhesion-free 3D matrix \cite{SHIH2010L29} and tortuous geometry \cite{Reversat2020}, 2D confinement \cite{RUPRECHT2015673, lammermann2008rapid}, and even free from confinement next to a non-adhesive substrate \cite{AOUN20201157} and in bulk fluids \cite{AOUN20201157, Bagchi2017, barry2010dictyostelium, oNeill2018}.

Extensive modeling efforts have been made to identify the minimal ingredients for cell motility.
Theoretical models predict spontaneous cell polarization and motility for 2D adhesion-mediated migration \cite{kozlov2007model, CallanJones2008, ziebert2012, doubrovinski2011, BlanchMercader2013, shao2010computational}.
For amoeboid migration, these processes are typically explained through myosin motor activity in models that neglect filament polymerization \cite{hawkins2011spontaneous, recho2013contraction, Voituriez2016, Farutin2019, Mietke2019b}.
However, the role of molecular motors in amoeboid migration has been challenged by experiments on T lymphocytes showing that retrograde flow is induced even in the absence of myosin activity \cite{AOUN20201157}. By contrast, inhibiting actin polymerization led to a significant decrease in migration. This suggests that actin polymerization is the predominant actor in amoeboid cell motility \cite{AOUN20201157}.
In line with these findings, a recent model neglecting myosin activity showed that a self-sustained, cell-scale retrograde flow can spontaneously emerge due to actin polymerization alone, even in the absence of cell shape changes \cite{Schmidt2025}.
This raises the question of whether the highly dynamic shape changes observed in motile amoeboid cells can arise from actin polymerization alone.

Here, we introduce a model for an amoeboid cell in a quasi-2D geometry, which corresponds to several experimental setups, such as migration between to planar plates \cite{Stankevicins2020, RUPRECHT2015673} or next to a non-adhesive substrate \cite{AOUN20201157}.
By analyzing the linear stability of the model, we show that the state of an unpolarized non-motile cell can spontaneously become unstable for a sufficiently large polymerization velocity of cortical actin filaments, leading to the onset of migration. For shape changing modes, stationary and Hopf bifurcations are found depending on parameters.
Numerical simulations of the full model reveal a rich variety of motile states, including migration along linear, circular, and zigzag trajectories, as well as irregular dynamics associated with complex shape deformations.
Furthermore, we find a non-motile regime characterized by steady, multi-lobed shapes.

This work is structured as follows.
We introduce our model in section \ref{sec:model}. We report on our results on the spontaneous onset of linear migration in section \ref{sec:Migration}.
We discuss circular migration in section \ref{sec:circular} followed by protrusive dynamics in section \ref{sec:protrusions}.
Finally, we conclude our findings in section \ref{sec:conclusion}.
\section{Model}
\label{sec:model}
\begin{figure}[h]
	\includegraphics[width=\columnwidth]{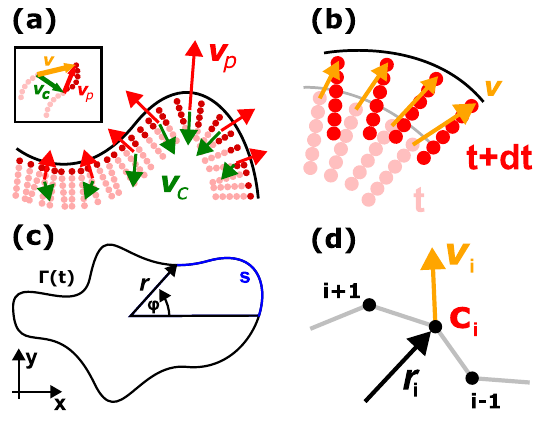}
	\caption{%
		(a) Cortex velocity $\bm v_c$ (green arrows) and polymerization velocity $\bm v_p$ (red arrows) along the cortex. Filaments are light-red rods and newly polymerized building blocks dark red circles, cell membrane is the black line.
		Inset: Full velocity (orange arrow) is sum of cortex velocity and polymerization velocity.
		(b) Full velocity $\bm v$ (orange arrows) evolves cell shape and advects actin filaments, with light-red rods representing filaments at time $t$ and red rods filaments at time $t + dt$.
		(c) Parametrization of the cell contour $\Gamma(t)$ using plane polar coordinates with vector $\bm r = r \bm e_r$ and angle $\varphi$. $s$ (blue line) shows contour line.
		(d) Numerical discretization of the cortex into nodes at positions $\bm r_i$ with actin filament concentration $c_i$ and full velocity $\bm v_i$.
		}
	\label{fig:sketch_model}
\end{figure}

We introduce a cross-sectional model for a single amoeboid cell in a planar geometry.
Our model belongs to the general class of active gel models \cite{Kruse2004}, which have been successfully used to describe myosin-driven contraction of the cortex and the resulting motility \cite{Voituriez2016,Farutin2019}.
We model the actin cortex as a thin, 1D shell of viscous compressible fluid that conforms to the membrane shape of the 2D cell. 
The cortex is represented as a closed curve $\Gamma (t)$ that moves in time $t$.
The position along the cortex is given by the vector $\bm r(\Gamma(t)) = (x,y)^T$.
We denote the cell perimeter as $p(\Gamma(t))$, the outwards-pointing unit normal vector by $\bm n (\bm r, t)$, and the tangent unit vector, pointing in positive direction along the curve, by $\bm \tau(\bm r, t)$.
The motion of actin filaments is given by two different velocity fields defined along the cell contour:
The cortex velocity $\bm v_c (\bm r, t)$ describes the deformation rate of the cortex. 
It accounts for the passive viscous response caused by the forces acting on the cortex as described below. 
The active growth of filaments due to polymerization is described by the polymerization velocity $\bm v_p (\bm r, t)$, see Fig.~\ref{fig:sketch_model} (a).
Their sum, 
\begin{equation}
	\label{eq_full_velocity}
	\bm v = \bm v_c +  \bm v_p
\end{equation}
is denoted as the full velocity which describes the motion of filament end points in the cell frame, see Fig.~\ref{fig:sketch_model} (b).
The full velocity determines the evolution of the cell boundary,
\begin{equation}\label{eq_full_velocity_cell_shape}
	\frac{\partial \bm r}{\partial t} = \bm v.
\end{equation}

The deformation rate of the cortex is governed by the force balance, given by
\begin{equation}\label{eqForceBalance}
	-\zeta \bm v_c + \bm f^\text{visc} (\bm v_c) + \bm f^\text{tens} + \bm f^\text{area} = \bm 0.
\end{equation}
The first term in Eq.~\eqref{eqForceBalance} is a drag force accounting for friction between the cortex and external substrate, with $\zeta$ being the friction coefficient.
The second term in Eq.~\eqref{eqForceBalance} denotes the viscous force. It is caused by viscous friction between filaments that make up the cortex and is given by
\begin{equation}\label{eqViscousForce}
	\bm f^\text{visc} (\bm v_c) = \bm \nabla_\Gamma \bm \cdot \mathbf{\sigma}_\Gamma (\bm v_c).
\end{equation}
Here, $\bm \nabla_\Gamma = \mathbb{I}_\Gamma \cdot \bm \nabla$ is the gradient taken along the contour. $\mathbb{I}_\Gamma = \mathbb{I} - \bm n \bm n$ is the  line projection operator and $\mathbb{I}$ the unit matrix.
$\mathbf{\sigma}_\Gamma$ is the passive viscous stress along the contour of the cell. 
It is given by
\begin{equation}\label{eqPassiveViscousStress}
	\mathbf{\sigma}_\Gamma (\bm v_c) = \eta \left( \bm \nabla_\Gamma \bm \cdot \bm v_c \right) \mathbb{I}_\Gamma
\end{equation}
where $\eta$ is the cortex viscosity.
The third contribution in Eq.~\eqref{eqForceBalance} is a force caused by the tension of the plasma membrane, given by
\begin{equation}\label{eqTensionForce}
	\bm f^\text{tens} = - \gamma H \frac{p_\text{max} - p_r}{p_\text{max} - p} \bm n .
\end{equation}
Here, $\gamma$ is the line tension, $H (\bm r, t)$ the membrane curvature, $p_\text{max}$ the maximum cell perimeter, and $p_r < p_\text{max}$ a reference perimeter. 
Eq.~\eqref{eqTensionForce} prevents large deviations of $p$ from its reference value.
Note that in the $p_\text{max} \rightarrow \infty$ limit, Eq.~\eqref{eqTensionForce} reduces to a classic Young-Laplace law.
The final contribution in Eq.~\eqref{eqForceBalance} is an area force, accounting for the conserved mass of the cytoplasm enclosed by the cell membrane. It is given by
\begin{equation}\label{eqAreaForce}
	\bm f^\text{area} = - \frac{2 \kappa}{A_0} \left( A - A_0 \right) \bm n,
\end{equation}
with area modulus $\kappa$, instantaneous cell area $A(t)$, and reference area $A_0$.
Eq.~\eqref{eqAreaForce} represents a force due to an effective difference between the hydrostatic pressures inside and outside the cell, $\Delta P$.

We denote by $c (\bm r, t)$ the actin concentration that defines the number of polymerizing filaments per unit contour length.
We assume that filament end points move with the full velocity $\bm v$ along the membrane.
For a contour line element $ds$, the filament number is given by $d \mathcal{N} = c \ ds$.
Without turnover or diffusion, $d \mathcal{N}$ is conserved in time in the filament co-moving frame, 
\begin{equation}\label{eq:filament_conservation}
	\frac{D (d\mathcal{N})}{Dt} = c \frac{D (ds)}{Dt} + ds \frac{Dc}{Dt} = 0
\end{equation}
where $\frac{D }{Dt} = \partial_t + \bm v \cdot \bm \nabla_\Gamma$ is the material derivative.
With the dilatation rate $\frac{1}{ds} \frac{D(ds)}{Dt} = \bm \nabla_\Gamma \cdot \bm v$, it follows from Eq.~\eqref{eq:filament_conservation}
$\frac{Dc}{Dt} = - c \bm \nabla_\Gamma \cdot \bm v$.
This means that in the filament frame, the concentration changes only due to local extension or contraction of the contour.
Together, the time-evolution of the actin concentration in the filament frame is given by
\begin{equation}
	\label{eq_adv_diff_filament_frame}
	\frac{Dc}{Dt} = - c \bm \nabla_\Gamma \cdot \bm v + D \Delta_\Gamma c + \beta (c_0 - c).
\end{equation}
Here, we include filament diffusion along the cortex with diffusivity $D$ and Laplace Beltrami operator \mbox{$\Delta_\Gamma = \bm \nabla_\Gamma \cdot \bm \nabla_\Gamma$}. 
$\beta$ is a restoration rate that tends to keep the concentration close to the homeostatic concentration $c_0$ and describes the rate at which filaments cease to polymerize, e.g., due to capping or depolymerization.

For simplicity,  we choose in our analytical analysis the cell frame as a reference frame where we track the concentration evolution at a given point of the cell contour. This results in an additional advection term  $- \bm v \cdot \bm \nabla_\Gamma c$ on the right-hand side of Eq.~\eqref{eq_adv_diff_filament_frame}, which becomes
\begin{equation}
	\label{eq_adv_diff}
	\frac{\partial c}{\partial t} = - \bm \nabla_\Gamma \cdot ( \bm v c) + D \Delta_\Gamma c + \beta (c_0 - c).
\end{equation}

Our model is closed by an expression for the polymerization velocity.
Filaments are assumed to polymerize normal to the membrane with velocity
\begin{align}
	\label{eq_vp}
	\bm v_p = V e^{- \frac{c}{c_r}} \bm n.
\end{align}
Herein, $V$ is the maximum polymerization speed and $c_r$ a reference concentration.
Eq.~\eqref{eq_vp} implies reduced polymerization in regions of elevated filament concentration, which is motivated by the experimental evidence that competition for monomeric building blocks limits filament growth \cite{Colin2023}.
For a more quantitative rationale of Eq.\ \eqref{eq_vp} we refer to previous work \cite{Schmidt2025}.

In the following we use as characteristic scales $R_0$ for length, $\beta^{-1}$ for time,  $c_r$ for concentration, and $\zeta R_0 \beta $ for force line density.
We rescale all variables but keep the same notations for simplicity.
With this the dimensionless model equations read
\begin{align}
	\label{eq_model_dimless1}
	\frac{\partial \bm r}{\partial t} = &\bm v_c + \bar{V} e^{-c} \bm n \\
	\label{eq_model_dimless2}
	\notag
	\bm 0 = &- \bm v_c + \bar{\eta} \bm \nabla_\Gamma \cdot \left[ \left( \bm \nabla_\Gamma \cdot \bm v_c \right) \mathbb{I}_\Gamma \right]\\ 
	&- \bar{\gamma} H \frac{p_\text{max} - p_r}{p_\text{max} - p} \bm n - 2 \bar{\kappa} \left( A - \frac{A_0}{R_0^2} \right) \\
	\label{eq_model_dimless3}
	\frac{\partial c}{\partial t} = & - \bm \nabla_\Gamma \cdot ( \bm v c) + \bar{D} \Delta_\Gamma c + \beta (\bar{c}_0 - c).
\end{align}
That is, the model is fully characterized by six independent dimensionless parameters,
\begin{align}\notag
	\bar{V} &= \frac{V}{R_0 \beta}, \bar{\eta} = \frac{\eta }{\zeta R_0^2}, \bar{\gamma} = \frac{\gamma}{\zeta \beta R_0^2},\\
	\bar{\kappa} &= \frac{\kappa R_0}{\zeta \beta A_0}, \bar{D} = \frac{D}{\beta R_0^2}, \bar{c}_0 = \frac{c_0}{c_r}.
\end{align}
\section{Spontaneous onset of motility leads to linear migration}
\label{sec:Migration}
In this section, we show that the base state of the model can become spontaneously unstable with respect to a front-back polarization of the cell, leading to migration without myosin motor activity or cell shape changes.
We present the results of the numerical simulations in section~\ref{sec:MigrationNumerics} and the linear stability analysis in section~\ref{sec:MigrationLinearStab}.
\subsection{Numerical results}
\label{sec:MigrationNumerics}
The model has a stationary base state characterized by a  circular cell shape of radius $r = 1$ and homogeneous actin concentration $c = \bar{c}_0$ along the cortex.
The associated polymerization velocity $\bm v_p^\text{base}$ is isotropic and points radially outwards.
It is fully canceled by the inwards-pointing cortex velocity $\bm v_c^\text{base}$, so that $\bm v^\text{base} = \bm 0$ (see Methods \ref{sec:methods}).

\begin{figure*}[t]
	\includegraphics[width=\textwidth]{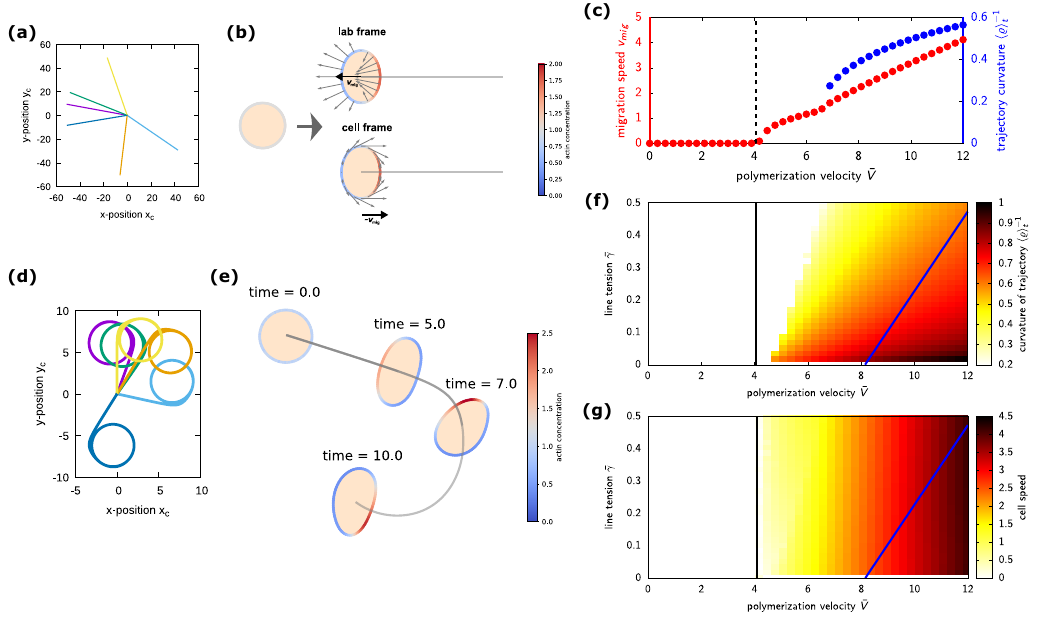}
	\caption{%
		\emph{Linear and circular migration.}
		(a): Linear cell trajectories for $\bar{V} = 6$.
		(b): Snapshots of the cell initially (left) and during linear migration with velocity $\bm v_\text{mig}$ (right) with trajectory shown in light gray.
		Gray arrows show full velocity in the laboratory (top) and cell frame (bottom). 
		Color code shows actin concentration (see legend).
		(c): Bifurcation diagram showing cell migration speed $v_\text{mig} = \abs{\bm v_\text{mig}}$ (red, left vertical axis) and trajectory curvature $\langle \varrho \rangle_t^{-1}$ (blue, right vertical axis) as function of polymerization velocity $\bar{V}$. 
		Black dashed line indicates the analytically calculated critical polymerization velocity.
		(d): Circular migration for $\bar{V} = 8$.
		(e): Time sequence of the cell (bold headlines show time) showing the transition from a resting state over straight motion to circular trajectory. 
		Color code shows actin concentration (see legend).
		(f) and (g) show phase diagrams with trajectory curvature $\langle \varrho \rangle_t^{-1}$  and migration speed $v_\text{mig}$ (color code, see legend) as a function of line tension $\bar{\gamma}$ and polymerization velocity $\bar{V}$. 
		Black (blue) line shows the analytically derived  threshold of the first (second) harmonic.
		}
	\label{fig:linearCircular}
\end{figure*}

We first focus on the limit of a vanishing viscous force ($\bar{\eta} = 0$). In this case, the tangential components of the forces in Eq.~\eqref{eqForceBalance} vanish and therefore all velocities become purely normal.
We find that for small polymerization velocities, the initial concentration perturbation decays for all modes and the base state is stable [see video 1]. 
For values of $\bar{V}$ above a critical value, the $l = 1$ concentration harmonic first grows and then saturates at a finite amplitude, which is accompanied by a translation of the cell. 
Hereby, filaments accumulate on one side of the cell, which becomes the cell rear, and deplete on the opposite side which turns into the cell front [see video 2].
Once the concentration distribution saturates, the cell attains a finite steady migration velocity $\bm v_\text{mig}$ and shape. 
The steady polarity leads to linear trajectories as shown in Fig.~\ref{fig:linearCircular}(a) where the migration direction is randomly determined during the initial spatial symmetry breaking.
Closely above the critical polymerization velocity, the cell shape is approximately circular, while at larger $\bar{V}$, it becomes increasingly elliptical with the major axis oriented perpendicular to the migration direction.

The migration mechanism is illustrated in Fig.~\ref{fig:linearCircular}(b), where we show the full velocity in the lab and the cell frame.
In the lab frame, the full velocity points in normal direction outwards in the low-concentration region and inwards in the high-concentration region, resulting in a displacement of the cell center relative to the external substrate.
In the cell frame, this velocity field becomes purely tangential and is directed from the cell front to the rear. This retrograde flow advects cortical filaments from the low concentration region to the high concentration region, further amplifying the concentration polarization via a positive feedback loop.

\subsection{Linear stability analysis}
\label{sec:MigrationLinearStab}
These numerical findings are quantitatively captured by our linear stability analysis of the first circular harmonic.
In linear approximation, this mode describes a rigid-body translation of the cell without shape change and therefore the tension force and area force are zero.
Substitution of Eq.~\eqref{eq:linear_ansatz_conc} for $l = 1$ into Eq.~\eqref{eq_vp} yields the linearized polymerization velocity 
\hbox{$\bm v_p = \bm v_p^\text{base} + \delta \bm v_{p, 1}$}
with the polymerization velocity perturbation
\begin{equation}\label{eqVp1Linear}
	\delta \bm v_{p, 1} = - \bar{V} e^{-\bar{c}_0} \delta C_1 \cos \varphi \bm e_r.
\end{equation}
$\delta \bm v_{p, 1}$ points radially outwards in regions of reduced concentration and radially inwards in regions of increased filament density, see also Fig.~\ref{fig:linearCircular}(b) (top).
This results in a shape-preserving displacement of the cell in the lab frame with  migration velocity
$\delta \bm v_\text{mig} = - \bar{V} e^{-\bar{c}_0} \delta C_1 \bm e_x$,
where $\delta \bm v_{c, 1} = - \delta \bm v_\text{mig}$.
The velocity in the cell frame is then given by the full velocity perturbation,
\begin{equation}\label{eqVFullFirstOrder}
	\delta \bm v_{1} = \delta \bm v_{c, 1} + \delta \bm v_{p, 1} = - \bar{V} e^{-\bar{c}_0} \delta C_1 \sin \varphi \ \bm e_\varphi.
\end{equation}
This expression describes a cell-scale, tangential flow of filaments from the low concentration to the high concentration region of the cortex, see also Fig.~\ref{fig:linearCircular}(b) (bottom) and therefore acts as a driver of the instability through a positive feedback loop.
Substitution of Eq.~\eqref{eqVFullFirstOrder} into Eq.~\eqref{eq_adv_diff} yields 
$\partial_t \delta {C}_1 = \lambda_1 \delta C_1$.
This means that the base state becomes unstable with respect to a perturbation of the first harmonic if the growth rate
\begin{equation}\label{eqLambda1}
	\lambda_1 = \bar{V} \bar{c}_0 e^{- \bar{c}_0} - \bar{D} - 1
\end{equation}
changes sign from negative to positive.
This is the case if $ \bar{V}$ exceeds a critical polymerization velocity $\bar{V}_c^{1}$.
Eq.~\eqref{eqLambda1} shows that actin polymerization is the driving mechanism of the instability, while tangential filament diffusion and restoration of the filament concentration counter-act it.
The analytical value for $\bar{V}_c^{1}$ obtained from Eq.~\eqref{eqLambda1} agrees well with the numerically found critical polymerization velocity at $\bar{V} \approx 4$ that marks the transition from the non-motile cell to a migratory state, see Fig.~\ref{fig:linearCircular}(c).
The growth rate in Eq.~\eqref{eqLambda1} is identical to the critical polymerization speed obtained for a spherical cell \cite{Schmidt2025} up to numerical prefactors.
\section{Secondary instability causes circular migration}
\label{sec:circular}
In this section, we show that for further increased polymerization velocities, the system undergoes a secondary bifurcation. It is characterized by a turning instability resulting in circular trajectories that take place already in the limit of vanishing inter-filament friction.
We report our numerical findings in section \ref{sec:CircularNumerics}, followed by a linear stability analysis for shape-changing modes in section \ref{sec:CircularLinearStab}.
\subsection{Numerical results}
\label{sec:CircularNumerics}
Exemplary trajectories for $\bar{V} = 8$ and $\bar{\eta} = 0$ are shown in Fig.~\ref{fig:linearCircular}(d).
The cell initially migrates along a straight path, then initiates a turning motion in a randomly chosen direction, followed by a closed circular trajectory with a constant radius [see also video 3].
Fig.~\ref{fig:linearCircular}(e) shows simulation snapshots for different times.
Starting from the non-polarized, circular state at $t = 0$, the cell transitions to straight motion accompanied by a front/back polarity and elliptical shape at $t = 5$.
Once a critical eccentricity is reached, the cell spontaneously breaks its left/right symmetry ($t = 7$) and the rearward, high-concentration region shifts laterally. 
This tangential motion causes the cell to change its migration direction towards the opposite side.
This behavior arises because high-concentration regions are attracted to regions of high contour curvature \cite{PhysRevE.95.012401}.
The final steady state of circular motion is characterized by an asymmetric shape and concentration profile ($t = 10$).

To characterize circular motion, we compute the radius of the cell trajectory $\varrho$.
In the following, we denote by $\langle \cdot \rangle_t$ the time average over a trajectory excluding the initial transient.
We approximate the trajectory curvature by $\langle \varrho \rangle_t^{-1}$.
As discussed above, the base state becomes unstable via a primary instability that leads to the onset of linear migration for $\bar{V} > \bar{V}_c^1$.
For further growing polymerization velocities, we observe a continuous increase of migration speed in Fig.~\ref{fig:linearCircular}(c).
The onset of circular motion takes place for $\bar{V}$ larger than a second critical value at $\bar{V} \approx 7$, where $\langle \varrho \rangle_t^{-1}$ attains a finite value. 
The curvature of trajectories then increases monotonically, that is, circles become smaller for further growing values of $\bar{V}$.
This shows that the onset of circular motion is a secondary bifurcation  since it requires that the cell has already undergone a breaking of front/back symmetry and attained a state of linear migration and elliptic shape deformation.

We now investigate the influence of line tension on cell migration.
Fig.~\ref{fig:linearCircular}(f) shows the trajectory curvature as a function of $\bar{\gamma}$ and $\bar{V}$. 
The critical polymerization velocity which marks the transition from linear to circular migration is shifted towards higher values of $\bar{V}$ for increasing $\bar{\gamma}$. 
This means that line tension, which suppresses cell shape changes, counter-acts the onset of circular motion.
Fig.~\ref{fig:linearCircular}(g) shows that the migration speed increases with growing $\bar{V}$ for all line tensions explored here.
Together, these results show that the onset of circular motion requires a sufficiently strong deviation from a circular cell shape. 
This is in contrast to the onset of linear migration described above which already takes place in the fixed-shape limit.
The influence of filament diffusivity on circular migration is shown in the Supplementary Information, section \ref{sec:circularDiffusivity}.
\subsection{Linear stability analysis}
\label{sec:CircularLinearStab}
Our numerical findings indicate a crucial role of cell shape changes for the onset of circular migration.
We therefore analyze the linear stability of shape changing modes, that is $l \geq 2$.
We find that the base state can become unstable with respect to a Hopf instability leading to oscillatory shape deformations and concentration dynamics.
The Hopf instability takes place for polymerization velocities above a critical value [see Methods section \ref{sec:linearStability} and Supplementary Information section \ref{sec:linearNoEta}]
\begin{equation}\label{eq:CriticalVHopf}
	\bar{V}_c^{l, \text{Hopf}} =  \frac{ e^{\bar{c}_0}}{\bar{c}_0} \left[ (l^2 -1) \bar{\gamma} + l^2 \bar{D} + 1 \right].
\end{equation}
This shows that  the instability is driven by normal flows due to actin polymerization and counter-acted by line tension, diffusion, and restoration of the concentration.
$\bar{V}_c^{l, \text{Hopf}}$ in Eq.~\eqref{eq:CriticalVHopf} increases with the mode index $l$ implying that higher harmonics require larger polymerization velocities to become unstable.
In Figs.~\ref{fig:linearCircular}(f) and \ref{fig:linearCircular}(g) we show the linear threshold in Eq.~\eqref{eq:CriticalVHopf} for the $l = 2$ harmonic (the most unstable shape-changing mode) as a function of the line tension.
For all parameters, the onset of circular motion takes place for polymerization velocities larger than the linear threshold for the $l =1$ and smaller than the one for the $l = 2$ mode.
An analytical determination of the threshold marking the onset of circular motion would require a nonlinear analysis which is outside the scope of this study.
\section{Protrusive dynamics for finite viscous force}
\label{sec:protrusions}
We now extend our analysis to the full model which includes the case of non-zero viscous friction between cortical filaments.
This leads to a finite viscous force in Eq.~\eqref{eqForceBalance} and therefore both normal and tangential flows are generally permitted.
Our numerical results in section \ref{sec:protrusionsNumerical} are complemented with a linear stability analysis in section \ref{sec:protrusionslinearStab}. 
\subsection{Numerical results}
\label{sec:protrusionsNumerical}
\begin{figure*}[t]
	\includegraphics[width=\textwidth]{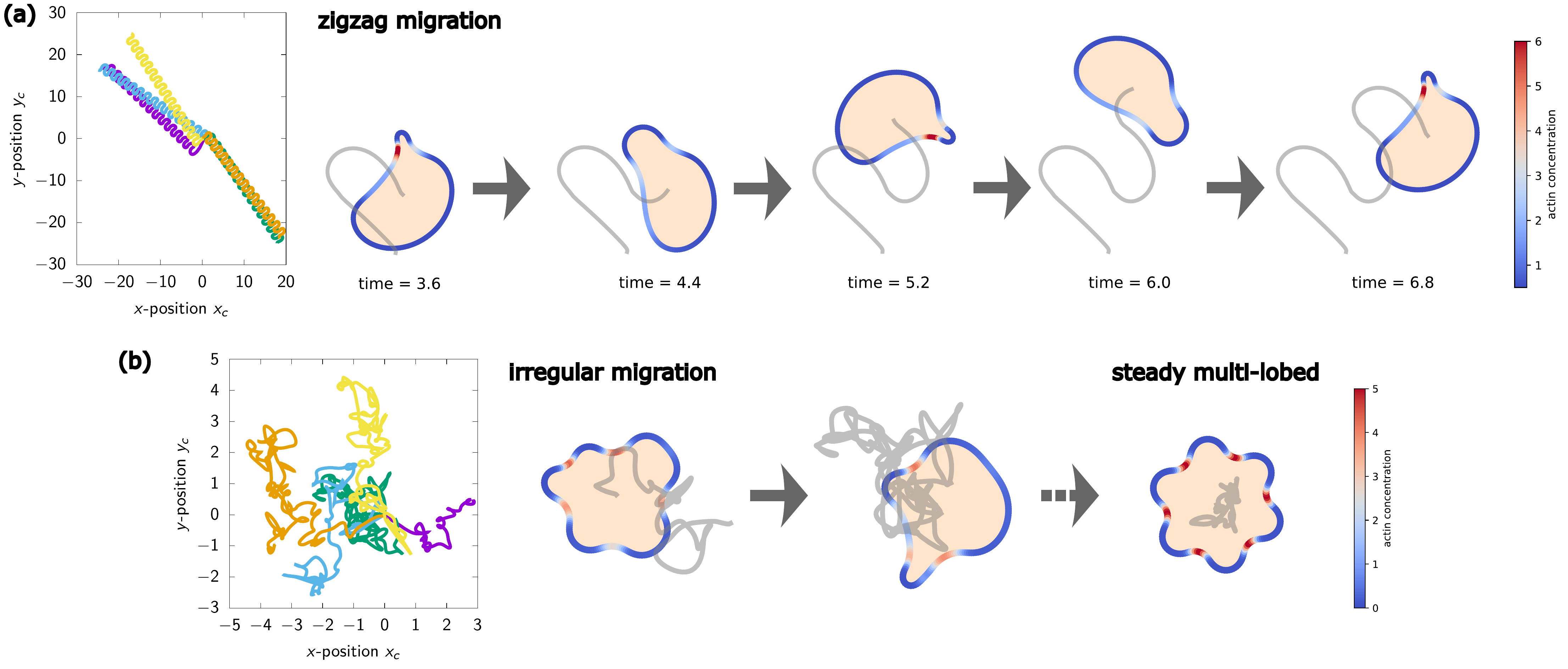}
	\caption{%
		\emph{Cell trajectories and shapes during protrusive dynamics.}
		(a) shows zigzag migration for $\bar{V} = 8.89$ and (b) irregular migration and steady multi-lobed shape for $\bar{V} = 12.44$.
		For each case, exemplary cell trajectories are shown on the left and time-series of cell shapes on the right with light-gray line representing cell trajectory up to given time instance (see bottom label) and color code showing actin filament concentration (see legend).
		Gray bold arrows indicate transition in time, gray dashed arrow possible transition that depends on parameters (see main text).
	}
	\label{fig:trajsZigzagIrregular}
\end{figure*}

In the following, we choose $\bar{D} = 0.1$, $\bar{\gamma} = 1$, and $\bar{\eta} = 0.03$ if not mentioned otherwise. 
For small and moderate polymerization velocities, we recover the above-described states of a non-motile cell, linear migration, and circular motion for increasing $\bar{V}$.
For further increasing polymerization velocities, the cell transitions to a zigzag-like trajectory, as shown in  Fig.~\ref{fig:trajsZigzagIrregular}(a) (left).
This migration type is characterized by alternating left and right turns which are accompanied by ample shape deformations [see video 4], see  Fig.~\ref{fig:trajsZigzagIrregular}(a) (right).
The zigzag trajectory arises from a complex periodic deformation pattern dominated by two lateral protrusions on opposite sides of the cell.
One protrusion grows to a cell-scale extension, pushing the cell boundary outwards. At the same time, the cortex retracts at the opposite side, causing the trailing protrusion to shrink into a smaller bulge.
Once a critical deformation is reached, the process reverses. 
During this cycle, actin filaments tend to accumulate at the cell rear and are transported dynamically along the cortex in an oscillatory pattern.
Because the deformation is asymmetric, the protrusion cycle results  not only in a left–right periodic motion but also a net forward displacement on average, creating a persistent, linear trajectory. 
This differs qualitatively from the circular paths described above in section \ref{sec:circular}, where the cell migrates around a fixed point without persistent orientation over time, causing a bounded trajectory.

\begin{figure*}[t]
	\includegraphics[width=\textwidth]{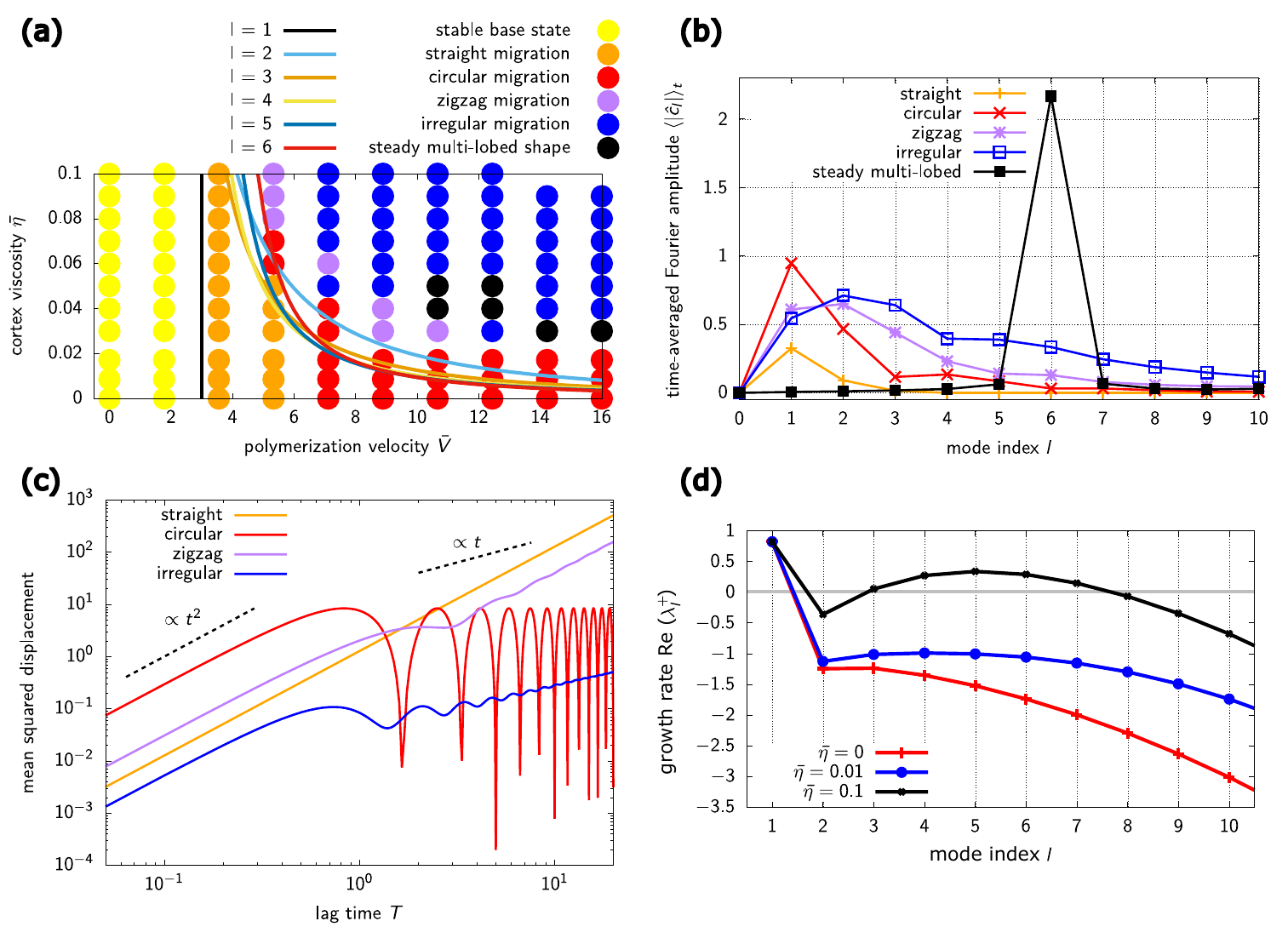}
	\caption{%
		\emph{Characterization of protrusive dynamics.}
		(a): State diagram showing different types of cell dynamics as a function of the dimensionless cortex viscosity $\bar{\eta}$ and polymerization speed $\bar{V}$. Colors of the dots correspond to different motion (see legend). 
		Lines indicate the analytically determined linear thresholds for the first six harmonics (see legend).
		(b): Time-averaged Fourier amplitude $\langle |\hat{c}_l| \rangle_t$, computed via Eq.~\eqref{eq:FourierTrafoConc}, as a function of the Fourier mode $l$ for different motility states (color code, see legend).
		(c): Mean squared displacement (MSD) [see Eq.~\eqref{eq:MSD}] as a function of lag time $T$ for various motile states of the cell (color code, see legend).
		Dashed lines indicate ballistic and diffusive behavior.
		(d): Linear dispersion relation showing growth rates $\text{Re}(\lambda_l^+)$ as a function of the mode index $l$ for different values of the dimensionless cortex viscosity $\bar{\eta}$. $\bar{V} = 5$, $\bar{\gamma} = 5$, $\bar{c}_0 = 1$, and $\bar{D} = 0.02$ are used.
	}
	\label{fig:quantificationProtrusive}
\end{figure*}

For further increasing values of $\bar{V}$, the cell undergoes yet another transition which results in irregular dynamics.
In this regime, we find highly dynamical states, which are characterized by protrusions of varying size that are continuously created and annihilated while the concentration is dynamically redistributed along the cortex in an irregular manner [see video 5], see  Fig.~\ref{fig:trajsZigzagIrregular}(b) (right).
Exemplary trajectories are shown in Fig.~\ref{fig:trajsZigzagIrregular}(b) (left).
We furthermore find that for some parameter values the cell ceases to migrate and adopts a fixed, multi-lobed shape, typically with five or six protrusions, see right-most panel of Fig.~\ref{fig:trajsZigzagIrregular}(b). The corresponding stationary concentration profile has pronounced local maxima in the concave regions and minima in convex regions of the cortex [see video 6].
This behavior is summarized in Fig.~\ref{fig:quantificationProtrusive} (a), where we show the various states as a function of the dimensionless cortex viscosity and polymerization velocity.
Irregular trajectories appear in the regime of large $\bar{V}$ and $\bar{\eta}$.
Zigzag trajectories are found in parameter ranges between straight or circular motion and irregular motion or steady multi-lobed states.

We characterize the various states by computing the Fourier transform of the actin concentration along the contour,
\begin{equation}\label{eq:FourierTrafoConc}
	\hat{c}_l (t) = \frac{1}{2 \pi} \int_{0}^{2 \pi} c (\varphi, t) e^{- i l \varphi} d \varphi.
\end{equation}
In Fig.~\ref{fig:quantificationProtrusive}(b) we plot the Fourier amplitude $\langle |\hat{c}_l| \rangle_t$ as a function of the mode index $l$. 
The amplitude spectrum exhibits qualitatively different behavior for the various motile states observed in our model.
For straight motion, the amplitude has a global maximum for $l=1$, a hallmark of persistent front/back polarity, whereas higher modes are barely present in the spectrum.
Circular migration is characterized by an increased concentration amplitude for the first harmonic and higher-order modes become more present compared to straight motion.
For zigzag and irregular motility, the maximum of the spectrum shifts towards larger modes. 
In these migration modes, persistent polarity breaks down and higher-order harmonics dominate.
The spectrum for the steady multi-lobed state is characterized by a marked single peak at the $l = 6$ mode which is caused by the observed six-fold symmetry of the stationary concentration profile.

We further quantify the trajectories of the various states using the mean squared displacement of the cell center
\begin{equation}\label{eq:MSD}
	\text{MSD} = \langle \left[ \bm r_c(t + T) - \bm r_c(t) \right]^2 \rangle_t
\end{equation}
with lag time $T$.
\begin{table}[h]
	\begin{tabular}{c  c  c}
		\hline
		\textbf{State:} & \textbf{Cortex viscosity $\bar{\eta}$:} & \textbf{Polym.\ velocity $\bar{V}$:} \\
		\hline
		straight migration & $0$ & $5.33$  \\
		circular migration & $0$ & $16.0$  \\
		zigzag migration & $0.03$ & $8.89$  \\
		irregular migration & $0.1$ & $8.89$  \\
		\hline
	\end{tabular}
	\caption{Parameters for simulations of protrusive cell dynamics.}
	\label{tab:parameters}
\end{table}
In Fig.~\ref{fig:quantificationProtrusive}(c) we plot the MSD for the different motile states with the parameters listed in tab.~\ref{tab:parameters}, including straight, circular, zigzag, and irregular migration, see also state diagram in Fig.~\ref{fig:quantificationProtrusive}(a).
For irregular trajectories we average the MSD over $50$ trajectories.
For all four motility patterns, the MSD exhibits ballistic growth at short lag times, indicating persistent directed motion, consistent with active self-propulsion.
The MSD changes for intermediate and long times, providing a unique signature for each type of motility.
For straight migration, we find classical ballistic behavior throughout all lag times.
In the case of zigzag motility, the MSD saturates for intermediate values of $T$. 
These lag times correspond to the time scale of periodic shape deformations that cause the cell to dynamically repolarize, leading to changes in its migration direction.
In the long time limit, the MSD grows quadratically in time. This indicates a second ballistic regime corresponding to the averaged linear migration for long times, see Fig.~\ref{fig:trajsZigzagIrregular}(a).
In the case of a circular trajectory, the MSD behaves oscillatory at intermediate and large lag times. The frequency of the oscillation is determined by the periodicity of the circular motion.
Once circular motion is attained, the MSD ceases to grow in time which is a consequence of the bounded trajectory.
For irregular motion, we observe oscillations of the MSD for intermediate lag times, albeit with a much smaller amplitude compared to circular motion, that stem from the highly dynamic shape deformations.
The oscillations decay for large lag times where the MSD grows linearly in time. This diffuse scaling in the long-time limit is a consequence of the polarity decorrelation and is in alignment with experimental measurements for amoeboid cells migrating in 2D confinement \cite{RUPRECHT2015673} and on non-adherent substrates \cite{AOUN20201157}.
\subsection{Linear stability analysis}
\label{sec:protrusionslinearStab}
These numerical findings highlight that a finite viscous force can give rise to much more complex dynamics compared to the $\bar{\eta} = 0$ limit.
To shed light on the mechanism that creates such high-wavenumber cortical protrusions, we analyze the linear stability of $l \geq 2$ modes for finite $\bar{\eta}$.
We first focus on the limit of a fixed cell shape for which we find a stationary instability that is caused to tangential cortical flows.
We find that the dynamics of the $l$-th concentration harmonic is governed by the linear growth rate [see Methods section \ref{sec:linearStability} and Supplementary Information section \ref{sec:linearFixedShape}]
\begin{equation}\label{eq:lambdaLStatFixedShape}
	\lambda_l = \bar{V} \bar{c}_0 e^{- \bar{c}_0} \left( 1 - \frac{A_3 \bar{V} \bar{c}_0 - A_4 \bar{\gamma}}{A_1 \bar{V} \bar{c}_0 - A_2 \bar{\gamma}} \right) - l^2 \bar{D} - 1
\end{equation}
with the constants $A_i$ given in Methods section \ref{sec:linearStability}.
$\lambda_l$ in Eq.~\eqref{eq:lambdaLStatFixedShape} is always real which means that the loss of stability of the homeostatic base state takes place via a stationary instability.
Eq.~\eqref{eq:lambdaLStatFixedShape} shows that high wavenumber modes can become unstable for sufficiently large polymerization velocities. 
Filament diffusion and restoration act against this tendency.
Note that normal flows are prohibited by the fixed-shape condition and the instability is solely caused by tangential flows due to viscous inter-filament friction.
If the fixed-shape assumption is relaxed, both stationary and Hopf instabilities are possible depending on the parameters, see Methods section \ref{sec:linearStability}.

The importance of the viscous force for the linear stability of the system is shown in Fig.~\ref{fig:quantificationProtrusive}(d).
For a cortex viscosity of zero, the growth rate decreases with growing mode index and the first harmonic is the only unstable mode.
For increasing $\bar{\eta}$, a second local maximum emerges in the dispersion relation for $l > 1$ with the growth rates of several higher-order harmonics becoming positive for $\bar{\eta} = 0.1$.
This means that these high-wavenumber modes, responsible for shape protrusions, become linearly unstable. 
In the nonlinear regime different modes couple which causes the complex dynamics observed in our simulations.
%
The linear thresholds for the first six modes are shown together with the numerical results in the state diagram in Fig.~\ref{fig:quantificationProtrusive}(a).
Zigzag trajectories and irregular dynamics are observed in the regime where the first circular harmonic and several shape-changing modes are linearly unstable.
\section{Conclusion}
\label{sec:conclusion}
In this work, we introduced and studied a 2D model for an amoeboid cell that takes into account cell shape changes driven by polymerization of cortical actin filaments.
We showed analytically and numerically that, already in the limit of a fixed cell shape, the state of an unpolarized cell becomes unstable above a critical polymerization velocity, leading to the spontaneous onset of front/back polarity and migration,  driven by cortical retrograde flow.
For further increasing polymerization velocities, we uncover the emergence of circular migration through a secondary bifurcation.
We report on a zigzag migration pattern, characterized by an asymmetric, periodic shape deformation, allowing the cell to persistently move at long times, and an irregular migration mode where highly dynamic protrusions grow and shrink erratically.
Moreover, we find non-motile stationary states that differ from the base state of the model in that they exhibit regular, pronounced shape protrusions and concentration variations along the cortex and occur after a series of transitions of motile states.
We characterized the appearance of these states as a function of the filament polymerization velocity and the ratio of cortical to substrate friction.
These findings are supported by linear stability analysis, revealing both a stationary instability and a Hopf bifurcation with distinct parameter regimes.

A similar motility cascade is caused for rigid autophoretic particles due to solute exchange with the surrounding bulk fluid \cite{hu2019chaotic, misbah2021universal, farutin2023motility}. In the present model, such trajectories arise from the coupling of shape and an active chemical species -- the polymerizing filaments -- that is confined to the domain of the cortex and arise already without interactions with a surrounding fluid.
The emergence of circular trajectories has been explained theoretically by a coupling of the cell shape and a polarity protein promoting membrane protrusions at the cell front \cite{PhysRevE.95.012401}, whereas in our model filaments accumulate in rearward regions.
Steady, turning, and zigzag motion have also been found in a model that predicts spontaneous onset of motility through mechanosensitive adhesion to the substrate \cite{PRXLife.1.023007}, which may not be present for all cell types.
Here, we find such dynamics arising from universal, polymerization-induced active stresses inside the cytoskeleton.

Our system exhibits dynamics similar to the recently discovered collective behavior of active particles that produces persistent and rotating worm-shaped swarms, caused by a competition between alignment and cohesion of self-propelling particles, and stationary asters where motility is lost as the system transitions to a more symmetric shape \cite{shea2025emergent}.

It has been shown that in a deterministic nonlinear system, chaotic transitions lead to diffusive scaling in the absence of noise \cite{geisel1982onset, hu2019chaotic}.
Our fully deterministic model exhibits above a critical control parameter a transition from regular to irregular dynamics which are characterized by a broad spatial concentration spectrum and  diffusive scaling in the long-time limit, therefore hinting to deterministic chaos.

The variety of states reported in our study are observed for several cell types in experiments.
Circular migration in a shape-preserving manner resembles migration of fish keratocytes \cite{ALLEN2020286}.
A persistent turning motion has also been observed for dendridic cells in 2D confinement \cite{Stankevicins2020} and for T lymphocytes in external shear flow after disruption of microtubules \cite{valignat2014lymphocytes}.
Migration in a zigzag pattern is documented for immune cells where it contributes to search efficiency \cite{li2015zigzag} and for a ``walking'' mode of \emph{Dictyostelium} amoebae on solids substrates \cite{van2011amoeboid} that is facilitated by pseudopods  growing on alternating sides of the cell.
The irregular migration mode that is characterized by highly dynamical shape changes shows ballistic motility for short times and becomes diffusive in the long-time limit.
These two regimes have also been identified for amoeboid cells migrating in 2D confinement \cite{RUPRECHT2015673} and on non-adherent substrates \cite{AOUN20201157}.
Remarkably, our model reproduces this behavior without relying on myosin contractiltiy.

Our findings highlight the importance of actin polymerization for amoeboid cell migration.
Notably, our model explains the spontaneous onset of cell polarization and protrusion-driven migratory states without relying on myosin motors while capturing the  short- and long-time scaling behavior.
Together, our findings mark a significant step towards a complete description of the mechanisms underlying amoeboid motility.
Our system shows similarity to aligning cohesive swarms of active particles \cite{shea2025emergent}, suggesting that transitions between motile states of broken symmetry and stationary, symmetric states are a shared feature of active matter systems in which shape and motion are coupled.
\section{Methods}
\label{sec:methods}
Our approach combines a linear stability analysis of the model, see in section \ref{sec:linearStability}, and a numerical solution of the full governing equations, as described in section \ref{sec:Numerics}.
\subsection{Linear stability analysis}
\label{sec:linearStability}
In our analytical description, we describe the cell contour using plane polar coordinates $(r, \varphi)$ where $r(\varphi, t) > 0$ is the radial distance from the cell center to the curve at angle $\varphi \in [0, 2 \pi )$.
The unit vectors along the radial and angular direction are given by $\bm e_r (\varphi) = \left( \cos \varphi, \sin \varphi \right)^T$ and $\bm e_\varphi (\varphi)  = \left( - \sin \varphi, \cos \varphi \right)^T$ and the position vector by $\bm r = r \bm e_r$. 
The arclength parameter $s \in [0, p]$ is related to the angle according to 
\begin{equation}\label{eqArclength}
	s(\varphi) = \int_{0}^{\varphi} \sqrt{r^2 + \left( \frac{dr}{d \varphi'} \right)^2} d \varphi'.
\end{equation}
The unit tangent is given by $\bm \tau = \frac{d \bm r}{d s}$ and the unit normal by $\bm n = \frac{d \bm \tau}{d s} / \abs{\frac{d \bm \tau}{d s}}$.

In the base state, the circular symmetry implies $\bm n = \bm e_r$ and $\bm \tau = \bm e_\varphi$.
To allow analytical treatment of the force balance, we replace the area force in Eq.~\eqref{eqForceBalance} by a normal pressure difference $\Delta P$.
The force balance in the base state then reads
\begin{equation}\label{eq_force_balance_base}
	- \bm v_c^\text{base} + \bm f^\text{visc}_\text{base} (\bm v_c^\text{base}) + \bm f^\text{tens}_\text{base} + \Delta P \bm e_r = \bm 0,
\end{equation}
where $\bm v_c^\text{base}$ is the cortex velocity associated with the base state.
From Eq.~\eqref{eq_vp} follows 
\begin{equation}\label{eq_vp_base}
	\bm v_p^\text{base} = \bar{V} e^{- \bar{c}_0} \bm e_r.
\end{equation}
The fixed shape implies that
\begin{equation}\label{eq_fixed_shape}
	\bm v \cdot \bm n = 0.
\end{equation}
From Eq.~\eqref{eq_full_velocity_cell_shape} then follows
\begin{equation}\label{eq_vc_base}
	\bm v_c^\text{base} = - \bar{V} e^{- \bar{c}_0} \bm e_r.
\end{equation}
Eq.~\eqref{eq_vc_base} describes the radial motion of material points of actin filaments towards the cell interior as new monomers are attached to the filaments close to the membrane.
The forces in Eq.~\eqref{eq_force_balance_base} are obtained as $\bm f^\text{tens}_\text{base} = - \bar{\gamma} \bm e_r$ and $\bm f^\text{visc}_\text{base} = \bar{\eta} \bar{V} e^{-\bar{c}_0} \bm e_r$.
Solving Eq.~\eqref{eq_force_balance_base} yields 
\begin{equation}\label{eq_force_balance_base0}
	\Delta P = \bar{\gamma} - \left( \bar{\eta} + 1 \right) \bar{V} e^{- \bar{c}_0}.
\end{equation}
This corresponds to the classical Laplace pressure due to membrane tension but diminished by a term proportional to the polymerization velocity $\bar{V}$ that stems from the dissipative forces caused by the radial inward cortex flow.

In order to analyze the linear stability of the base state, we decompose the shape and concentration into Fourier harmonics,
\begin{align}
	\label{eq:linear_ansatz_shape}
	r &= 1 + \delta R_l \cos (l \varphi)\\
	\label{eq:linear_ansatz_conc}
	c &= \bar{c}_0 + \delta C_l \cos (l \varphi)
\end{align}
with the dimensionless perturbation of the cell shape $\delta R_l (t)$ and actin concentration $\delta C_l (t)$.
We expand our model equations up to linear order in $\delta R_l (t)$ and $\delta C_l (t)$ for a given Fourier mode $l$.
In the linear regime, modes decouple and can be treated independently; furthermore sine and cosine modes decouple, so we omit the former.
The linearized normal and tangential unit vectors are obtained as
\begin{align}
	\label{eq:unit_normal_linear}
	\bm n &= \bm e_r + \delta R_l l \sin(l \varphi) \bm e_\varphi, \\
	\label{eq:unit_tangent_linear}
	\bm \tau &= \bm e_\varphi - \delta R_l l \sin(l \varphi) \bm e_r,
\end{align}
i.e., they contain first-order corrections along the directions perpendicular to the zeroth-order contribution.
We find  that the dynamics of shape and concentration can be described by the coupled system [see Supplementary Information section \ref{sec:linearFull}]
\begin{equation}\label{eq_shape_conc_coupled}
	\frac{\partial}{\partial t} \begin{pmatrix}
		\delta R_l \\
		\delta C_l
	\end{pmatrix} = \mathbf{J} \begin{pmatrix}
		\delta R_l \\
		\delta C_l
	\end{pmatrix}
\end{equation}
with the Jacoby matrix
\begin{align}\label{eq_Jacobi_full}
	\mathbf{J} = \begin{pmatrix}
		A_1 \bar{V}  e^{- \bar{c}_0}  - A_2 \bar{\gamma}  & - \bar{V} e^{-\bar{c}_0}  \\
		- A_3  \bar{V} e^{- \bar{c}_0} \bar{c}_0 + A_4  \bar{\gamma} \bar{c}_0 &  \bar{V} e^{- \bar{c}_0} \bar{c}_0 - l^2 \bar{D} - 1
	\end{pmatrix}.
\end{align}
The constants $A_i (\bar{\eta}, l) \geq 0$ ($i = 1, 2, 3,4$) in Eq.~\eqref{eq_Jacobi_full} are given by
\begin{align}\label{eq_constants_jacoby}
	A_1 &= \frac{(l^2 - 1) \bar{\eta} ( 2 + l^2 \bar{\eta})}{1 + (l^2 + 1) \bar{\eta}} \\
	A_2 &= \frac{(l^2 - 1) ( 1 + l^2 \bar{\eta})}{1 + (l^2 + 1) \bar{\eta}} \\
	A_3 &= \frac{(l^2 - 1) (l^2 + 2) \bar{\eta}}{1 + (l^2 + 1) \bar{\eta}} \\
	A_4 &= \frac{(l^2 - 1)}{1 + (l^2 + 1) \bar{\eta}}.
\end{align}
The system \eqref{eq_shape_conc_coupled} has a trivial solution given by $\delta R_l = 0$ and $\delta C_l = 0$ which corresponds to the isotropic base state as described in the main text.
The non-trivial solutions of \eqref{eq_shape_conc_coupled} are linear combinations of exponential functions with the exponents given by the eigenvalues of the Jacobian which are the roots of the characteristic polynomial
\begin{equation}\label{eq:characteristicPol}
	det \left( \textbf{J} - \lambda \textbf{I} \right) = 0.
\end{equation}
The solution of \eqref{eq:characteristicPol} is given by
\begin{equation}\label{eqEigenValuesJacobi}
	\lambda_l^\pm =  \frac{tr (\textbf{J})}{2}  \pm \frac{1}{2} \sqrt{ tr (\textbf{J})^2 - 4 det (\textbf{J})},
\end{equation}
where 
\begin{equation}\label{eq_trace_Jacobi}
	tr (\textbf{J}) = \bar{V}  e^{- \bar{c}_0} \left( A_1 + \bar{c}_0 \right)   - A_2 \bar{\gamma}  - l^2 \bar{D} - 1
\end{equation}
is the trace of the Jacobian and 
\begin{widetext}
\begin{equation}\label{eq_determinant_Jacobi}
	det (\textbf{J}) = \left(\bar{V} e^{- \bar{c}_0} \right)^2 \bar{c}_0 \left( A_1 - A_3 \right) + \bar{V} e^{- \bar{c}_0} \left[ \bar{\gamma} \bar{c}_0 \left( - A_2 + A_4 \right) - A_1 \left( l^2 \bar{D} + 1 \right) \right] + A_2 \bar{\gamma} \left( l^2 \bar{D} + 1 \right)
\end{equation}
\end{widetext}
its determinant.
The base state is linearly stable with respect to a perturbation of the $l$-th harmonic if the real parts of both growth rates $\lambda_l^\pm$ in Eq.~\eqref{eqEigenValuesJacobi} are negative.
This is the case if $tr (\textbf{J}) < 0$ and $det (\textbf{J}) > 0$.
Depending on the sign of the trace and determinant, the system gives rise to two different instabilities:
\begin{itemize}
	\item If $det (\textbf{J})$ in Eq.~\eqref{eq_determinant_Jacobi} changes sign from positive to negative, one of the eigenvalues in Eq.~\eqref{eqEigenValuesJacobi} becomes real and positive. The system then undergoes a stationary instability above a critical polymerization velocity $\bar{V}_c^{l, \text{Stat}}$ which follows directly from the $det (\textbf{J}) < 0$ condition.
	\item If $tr (\textbf{J})$ in Eq.~\eqref{eq_trace_Jacobi} changes sign from negative to positive while $det (\textbf{J})$ remains positive, Eq.~\eqref{eqEigenValuesJacobi} yields a pair of complex-conjugate eigenvalues with non-zero imaginary parts while the real part changes sign from negative to positive. This leads to a Hopf bifurcation above a critical polymerization velocity $\bar{V}_c^{l, \text{Hopf}}$ that follows directly from the $tr (\textbf{J}) > 0$ condition.
\end{itemize}
\subsection{Numerical implementation}
\label{sec:Numerics}
Numerically, we solve the governing equations on a freely moving and deforming cell cortex using the weak formulation, as detailed in Supplementary Information  section \ref{sec:method}.
We choose a discrete representation of the cortex using $N$ nodes forming a closed, 1D chain along the cell boundary.
We assign a position $\bm r_i = (x_i, y_i)^T$, velocity $\bm v_{i} = (v^x_{i}, v^y_{i})^T$, and actin filament concentration $c_i$ to each node $i$ with $i = 1, 2, ..., N$, see Fig.~\ref{fig:sketch_model} (d).
In each time iteration, we first compute the cell geometry and forces.
We then solve the force balance Eq.~\eqref{eq_model_dimless2} to obtain the cortex velocity. 
This gives access to the full velocity which allows to update the cell shape according to Eq.~\eqref{eq_model_dimless1}.
The concentration is evolved upon computing the advection diffusion Eq.~\eqref{eq_adv_diff_filament_frame}.
To ensure the stability of the mesh, we employ a tangential advection scheme of the nodes.
We validate each part of our algorithm separately by a comparison to analytically obtained solutions and find systematic convergence for different number of mesh nodes.

We initially set the cell shape to a circle with $r = 1$. 
The actin concentration is initialized to $c_0$ with an additional perturbation for all modes with small, random amplitudes.
If not mentioned otherwise, we choose as parameters $\bar{c}_0 = 1$, $\bar{D} = 0.5$, $\bar{\gamma} = 0.5$, $p_\text{max} = 2.5 \pi$, and $p_r = 2 \pi$.
\section{Acknowledgments}
\label{sec:Acknowledgments}
We thank P.~Recho for stimulating discussions.
We are grateful to CNES (Centre National d’Etudes Spatiales) and the French-German University (UFA/DFH) program “Living Fluids” (Grant No. CDFA-01-14) for financial support.

\clearpage
\appendix
\onecolumngrid

\renewcommand{\thefigure}{S\arabic{figure}}
\renewcommand{\thetable}{S\arabic{table}}
\renewcommand{\theequation}{S\arabic{equation}}

\setcounter{figure}{0}
\setcounter{table}{0}
\setcounter{equation}{0}

\section*{Supplementary Information}
\label{sec:SI}
\subsection{Linear stability analysis for shape deformation modes}
\label{sec:linearFull}
We give here the full calculation for the linear stability analysis of the $l \geq 2$ Fourier harmonics.
We first solve the passive, mechanical part of the problem in section \ref{sec:LinearForcebalance} by expanding the forces in Eq.~\eqref{eq_force_balance_linear} up to linear order.
From this we obtain the linearized cortex velocity.
We then expand the dynamical equations which involve the actin concentration in section \ref{sec:linearConcentrationDynamics}. 
Together, we obtain a coupled linear system of differential equations.
\subsubsection{Force balance and cortex velocity solution}
\label{sec:LinearForcebalance}
The force balance Eq.~\eqref{eqForceBalance} reads in linear approximation
\begin{equation}\label{eq_force_balance_linear}
	\bm f^\text{drag}_l (\bm v_{c,l}) + \bm f^\text{visc}_l (\bm v_{c,l}) + \bm f^\text{tens}_l + \Delta P \bm n = \bm 0
\end{equation}
where $\bm f^\text{drag}_l$, $\bm f^\text{visc}_l$, and $\bm f^\text{tens}_l$ are the linearized drag force, viscous force, and tension force, respectively, and $\bm v_{c,l}$ the linearized cortex velocity, which are calculated in the following.

We first compute the tension force. 
In general, the membrane curvature $H (\varphi, t)$ is given by
\begin{equation}\label{eq_mean_curv_general}
	H = \frac{r^2 + 2 \left(\frac{\partial r}{\partial \varphi} \right)^2 - r \frac{\partial^2 r}{\partial \varphi^2}}{\left[ r^2 + \left(\frac{\partial r}{\partial \varphi} \right)^2 \right]^\frac{3}{2}}.
\end{equation}
In linear approximation we obtain for the $l$-th order curvature \cite{PhysRevE.101.022404, CallanJones2008}
\begin{equation}\label{eqCurvatureLinear}
	H_l (\varphi, t) = 1 + \delta H_l (\varphi, t) = 1 + \delta R (t) (l^2 - 1)  \cos (l \varphi).
\end{equation}
Using Eqs.~\eqref{eq:unit_normal_linear} and \eqref{eq:unit_tangent_linear}, the tension force in Eq.~\eqref{eqTensionForce} yields up to linear order
\begin{equation}\label{eq_fTens_linear}
	\bm f^\text{tens}_l (\varphi, t) = \bm f_\text{base}^\text{tens} (\varphi) + \delta \bm f_l^\text{tens} (\varphi, t)
\end{equation}
with the tension force of the base state $\bm f_\text{base}^\text{tens} (\varphi) = - \bar{\gamma} \bm e_r$ and the $l$-th order tension force perturbation
\begin{equation}\label{eq_fTens_l}
	\delta \bm f_l^\text{tens} (\varphi, t) = - \bar{\gamma} \delta R_l(t)  \left[ (l^2 - 1)  \cos(l \varphi) \bm e_r (\varphi) + l \sin(l \varphi) \bm e_\varphi (\varphi) \right] .
\end{equation}
We now turn to the pressure term in \eqref{eq_force_balance_linear}.
Note that linear perturbations of the cell shape with $l \geq 1$ do not affect the cell area and therefore do not contribute to the pressure difference $\Delta P$.
Using Eq.~\eqref{eq:unit_normal_linear}, the pressure term in Eq.~\eqref{eq_force_balance_linear} yields
\begin{align}\label{eq_Deltap_linear}
	\Delta P \bm n (\varphi, t) = \Delta P \left[  \bm e_r (\varphi) + \delta R_l(t)  l \sin(l \varphi) \bm e_\varphi (\varphi) \right]
\end{align}
with $\Delta P$ given by Eq.~\eqref{eq_force_balance_base0}.

The dissipative forces in Eq.~\eqref{eq_force_balance_linear} depend on the cortex velocity.
We choose the ansatz
\begin{equation}\label{eq_velCortex_linear}
	\bm v_{c,l} (\varphi, t) = \bm v_c^\text{base} (\varphi) + \delta \bm v_{c, l} (\varphi, t).
\end{equation}
Herein, $\bm v_c^\text{base} (\varphi)$ is the cortex velocity of the base state given by Eq.~\eqref{eq_vc_base} and 
\begin{align}\label{eqAnsatyVelCortexL}
	\notag
	\delta \bm v_{c, l} (\varphi, t) &= \left[ \delta v_{r, l}^\text{cos} (t) \cos(l \varphi) + \delta v_{r, l}^\text{sin} (t) \sin(l \varphi) \right]  \bm e_r (\varphi)\\
	&+ \left[ \delta v_{\varphi, l}^\text{cos} (t) \cos(l \varphi) + \delta v_{\varphi, l}^\text{sin} (t) \sin(l \varphi) \right]  \bm e_\varphi (\varphi)
\end{align}
$l$-th order cortex velocity perturbation. The amplitudes $\delta v_{r, l}^\text{cos} (t)$, $\delta v_{r, l}^\text{sin} (t)$, $\delta v_{\varphi, l}^\text{cos} (t)$, and $\delta v_{\varphi, l}^\text{sin} (t)$ are determined in the following.
Together, we obtain from Eq.~\eqref{eqViscousForce} the linearized $l$-th order viscous force
\begin{equation}\label{eqViscousForceLinear}
	\bm f_l^\text{visc} (\varphi, t) = \bm f_\text{base}^\text{visc} (\varphi) + \delta \bm f_l^\text{visc} (\varphi, t)
\end{equation}
with the contribution from the base state $\bm f^\text{visc}_\text{base} = \bar{\eta} \bar{V} e^{-\bar{c}_0} \bm e_r$ and the $l$-th order viscous force perturbation
\begin{align}\label{eqViscousForcePerturbation}
	\delta \bm f_l^\text{visc}  (\varphi, t) = \delta f^\text{visc}_{l, r} (\varphi, t)  \bm e_r (\varphi) + \delta f^\text{visc}_{l, \varphi} (\varphi, t)  \bm e_\varphi (\varphi)
\end{align}
with
\begin{align}\label{eqViscousForceLinearCoeffs}
	\delta f^\text{visc}_{l, r} (\varphi, t) &= \bar{\eta} \left[ - \left( l \delta v_{\varphi, l}^\text{sin} + \delta v_{r, l}^\text{cos} - (l^2 - 2) \delta R_l \bar{V} e^{- \bar{c}_0} \right) \cos (l \varphi) + \left( l \delta v_{\varphi, l}^\text{cos} - \delta v_{r, l}^\text{sin} \right) \sin (l \varphi) \right], \\
	\delta f^\text{visc}_{l, \varphi} (\varphi, t) &= l \bar{\eta} \left[ \left( - l \delta v_{\varphi, l}^\text{cos} + \delta v_{r, l}^\text{sin} \right) \cos (l \varphi) - \left( l \delta v_{\varphi, l}^\text{sin} + \delta v_{r, l}^\text{cos} \right) \sin (l \varphi) \right].
\end{align}

Using equations
\eqref{eq_force_balance_base0},
\eqref{eq_fTens_l},
\eqref{eq_Deltap_linear},
\eqref{eqAnsatyVelCortexL},
\eqref{eqViscousForceLinearCoeffs},
we can write down the force balance equation in $l$-order which yields
\begin{align}
	\label{eq:force_balance_l_radial}
	\notag
	&- \delta v_{r, l}^\text{cos} (t) \cos(l \varphi) - \delta v_{r, l}^\text{sin} (t) \sin(l \varphi) \\
	&+ \bar{\eta} \left[ - \left( l \delta v_{\varphi, l}^\text{sin} + \delta v_{r, l}^\text{cos} - (l^2 - 2) \delta R_l \bar{V} e^{- \bar{c}_0} \right) \cos (l \varphi) + \left( l \delta v_{\varphi, l}^\text{cos} - \delta v_{r, l}^\text{sin} \right) \sin (l \varphi) \right]
	- \bar{\gamma} \delta R_l (l^2 - 1)  \cos(l \varphi) = 0
\end{align}
for the radial component and
\begin{align}
	\label{eq:force_balance_l_angular}
	\notag &- \delta v_{\varphi, l}^\text{cos} (t) \cos(l \varphi) - \delta v_{\varphi, l}^\text{sin} (t) \sin(l \varphi) \\
	&+ l \bar{\eta} \left[ \left( - l \delta v_{\varphi, l}^\text{cos} + \delta v_{r, l}^\text{sin} \right) \cos (l \varphi) - \left( l \delta v_{\varphi, l}^\text{sin} + \delta v_{r, l}^\text{cos} \right) \sin (l \varphi) \right]
	- \left( \bar{\eta} + 1 \right) \bar{V} e^{- \bar{c}_0} l \delta R_l \sin(l \varphi) = 0
\end{align}
for the angular component.
Upon projection of Eqs.~\eqref{eq:force_balance_l_radial} and \eqref{eq:force_balance_l_angular} on sine and cosine harmonics we solve for the cortex velocity coefficients, which are obtained as
\begin{align}\label{eqSolCoeffsVelCortexL}
	\notag
	\delta v_{r, l}^\text{cos} &= (l^2 - 1) \delta R_l(t) \frac{\bar{\eta} \bar{V}  e^{- \bar{c}_0} (2 + l^2 \bar{\eta}) - \bar{\gamma} (1 + l^2 \bar{\eta})}{1 + (l^2 + 1) \bar{\eta}} , \\
	\notag
	\delta v_{r, l}^\text{sin} &= 0, \\
	\notag
	\delta v_{\varphi, l}^\text{cos} &= 0, \\
	\delta v_{\varphi, l}^\text{sin} &=  l \delta R_l(t) \frac{(l^2 - 1) \bar{\eta} \bar{\gamma} - \bar{V} e^{- \bar{c}_0} \left[ 1 + 2 \bar{\eta} + (l^2 - 1) \bar{\eta}^2 \right]}{1 + (l^2 + 1) \bar{\eta}} .
\end{align}
\subsubsection{Polymerization velocity and shape and concentration evolution}
\label{sec:linearConcentrationDynamics}
In order to obtain the full velocity, which describes the cell shape evolution and advection of filaments along the contour, we need to compute the polymerization velocity in linear approximation. 
From Eq.~\eqref{eq_vp} follows with Eqs.~\eqref{eq:linear_ansatz_conc} and \eqref{eq:unit_normal_linear}
\begin{equation}\label{eq_vp_l_linearized_dimless}
	\bm v_p (\varphi, t) = \bm v_{p}^\text{base} (\varphi) + \delta \bm v_{p,l} (\varphi, t) + \mathcal{O} (\delta R_l \delta C_l).
\end{equation}
with the base state contribution $\bm v_{p}^\text{base} (\varphi)$ given by Eq.~\eqref{eq_vp_base}
and the polymerization velocity perturbation
\begin{equation}\label{eqVplLinear}
	\delta \bm v_{p,l} (\varphi, t) = \bar{V} e^{-\bar{c}_0} \left[ - \delta C_l(t) \cos(l \varphi) \bm e_r (\varphi) + \delta R_l(t) l \sin(l \varphi) \bm e_\varphi (\varphi) \right].
\end{equation}
Together, we obtain from Eq.~\eqref{eqAnsatyVelCortexL} with the coefficients \eqref{eqSolCoeffsVelCortexL} and Eq.~\eqref{eqVplLinear} the full velocity perturbation
\begin{align}\label{eq_deltaVL}
	\notag
	\delta \bm v_{l} (\varphi, t) &= \delta \bm v_{c, l} (\varphi, t) + \delta \bm v_{p, l} (\varphi, t) \\
	\notag
	&= \left[ (l^2 - 1) \frac{\bar{\eta} \bar{V}  e^{- \bar{c}_0} (2 + l^2 \bar{\eta}) - \bar{\gamma} (1 + l^2 \bar{\eta})}{1 + (l^2 + 1) \bar{\eta}} \delta R_l(t) - \bar{V} e^{-\bar{c}_0} \delta C_l(t) \right] \cos(l \varphi) \bm e_r (\varphi) \\
	&+ \left[ \frac{(l^2 - 1) \bar{\eta} \bar{\gamma} - \bar{V} e^{- \bar{c}_0} \left[ 1 + 2 \bar{\eta} + (l^2 - 1) \bar{\eta}^2 \right]}{1 + (l^2 + 1) \bar{\eta}} + \bar{V} e^{-\bar{c}_0}  \right] \delta R_l(t) l  \sin(l \varphi)  \bm e_\varphi (\varphi).
\end{align}
The cell shape evolves in time according to
\begin{align}\label{eq_shape_evolution}
	\frac{\partial \bm r}{\partial t} \cdot \bm n \stackrel{\eqref{eq_full_velocity_cell_shape}}{=} \left( \bm v_{c,l} + \bm v_{p,l} \right) \cdot \bm n \stackrel{\eqref{eq:unit_normal_linear}}{=} \delta \bm v_{l} \cdot \bm e_r.
\end{align}
Note that $\bm v_\text{base} = \bm 0$ in Eq.~\eqref{eq_shape_evolution}.
Taking the time-derivative of Eq.~\eqref{eq:linear_ansatz_shape} yields
\begin{equation}\label{eq:timeEvolutionShapeLinear}
	\frac{\partial \bm r}{\partial t} (\varphi, t) = \frac{\partial \delta R_l}{\partial t}  (t) \cos (l \varphi) \bm e_r.
\end{equation}
Equating \eqref{eq_shape_evolution} and \eqref{eq:timeEvolutionShapeLinear} and using \eqref{eq_deltaVL} yields upon projection onto the cosine mode
\begin{equation}\label{eq_rhoDot}
	\frac{\partial \delta R_l}{\partial t}  (t) = (l^2 - 1) \frac{\bar{\eta} \bar{V}  e^{- \bar{c}_0} (2 + l^2 \bar{\eta}) - \bar{\gamma} (1 + l^2 \bar{\eta})}{1 + (l^2 + 1) \bar{\eta}} \delta R(t) - \bar{V} e^{-\bar{c}_0} \delta C_l(t).
\end{equation}

Using the full velocity perturbation in Eq.~\eqref{eq_deltaVL} we can compute the advection term in Eq.~\eqref{eq_adv_diff}, which yields
\begin{align}\label{eq_advection_linear}
	\bm \nabla^l \cdot \left[ c \delta \bm v_l  \right] \approx \bar{c}_0  \bm \nabla^l \cdot \delta \bm v_l = \left[  \frac{\bar{\gamma} (l^2 - 1) - (l - 1)(l + 1) (l^2 + 2) \bar{\eta} \bar{V} e^{- \bar{c}_0}}{1 + (l^2 + 1) \bar{\eta}} \delta R_l (t) + \bar{V} e^{- \bar{c}_0} \delta C_l (t) \right] \bar{c}_0  \cos(l \varphi).
\end{align}
The full concentration evolution is obtained as
\begin{align}\label{eq_cdot_linear}
	\frac{\partial \delta C_l}{\partial t}  (t) = \bar{c}_0  \frac{\bar{\gamma} (l^2 - 1) - (l - 1)(l + 1) (l^2 + 2) \bar{\eta} \bar{V} e^{- \bar{c}_0}}{1 + (l^2 + 1) \bar{\eta}} \delta R_l (t) + \left( \bar{c}_0 \bar{V} e^{- \bar{c}_0} - l^2 \bar{D} - 1 \right) \delta C_l (t).
\end{align}
Eqs.~\eqref{eq_rhoDot} and \eqref{eq_cdot_linear} form a linear system of coupled differential equations which can be written as Eq.~\eqref{eq_shape_conc_coupled} in the main text with the Jacoby matrix given by Eq.~\eqref{eq_Jacobi_full} in the main text.
\subsubsection{Special case of a fixed cell shape}
\label{sec:linearFixedShape}
We now want to focus on the limit case of a fixed cell shape.
In this case, no normal flows exist according to Eq.~\eqref{eq_full_velocity_cell_shape} which translates in the linear regime to the limit of $\partial_t {\delta R}_l \rightarrow 0$.
With this we can solve Eq.~\eqref{eq_rhoDot} for the shape coefficient, yielding
\begin{equation}\label{eqShapeCoefficientLinear}
	\delta R_l = \frac{\bar{V} e^{- \bar{c}_0} \delta C_l}{A_1 \bar{V} e^{- \bar{c}_0} - A_2 \bar{\gamma}} .
\end{equation}
Substitution into Eq.~\eqref{eq_shape_conc_coupled} yields a single equation which describes the dynamics of the concentration perturbation,
\begin{equation}
	 \frac{\partial {\delta C}_l}{\partial t}  = \lambda_l \delta C_l.
\end{equation}
Herein,
\begin{equation}\label{eq:lambdaLFixedShape}
	\lambda_l = \bar{V} \bar{c}_0 e^{- \bar{c}_0} \left[ 1 - \frac{A_3 \bar{V} \bar{c}_0 - A_4 \bar{\gamma}}{A_1 \bar{V} \bar{c}_0 - A_2 \bar{\gamma}} \right] - l^2 \bar{D} - 1.
\end{equation}
is the linear growth rate of the $l$-th harmonic.
Note that $\lambda_l$ in Eq.~\eqref{eq:lambdaLFixedShape} is always real which precludes an oscillatory instability.
The base state becomes unstable via a stationary instability above a critical polymerization velocity which follows from the $\lambda_l > 0$ condition.

Eq.~\eqref{eq:lambdaLFixedShape} highlights that actin polymerization is the mechanism which drives the instability whereas filament diffusion and restoration counter-act it.
It is easy to see that the origin of this instability is the finite viscous friction inside the cortex: Consider the case of $\bar{\eta} = 0$ for which the constants in Eq.~\eqref{eq_constants_jacoby} simplify to $A_1 = 0$, $A_2 = l^2 - 1$, $A_3 = 0$, $A_4 = l^2 - 1$ and Eq.~\eqref{eq:lambdaLFixedShape} yields $\lambda_l = - l^2 \bar{D} - 1$. That is, $\lambda_l$ is always negative, independently of the polymerization velocity, and therefore the base state cannot become unstable.
\subsubsection{Special case of vanishing cortex viscosity}
\label{sec:linearNoEta}
A second insightful special case of the full linear system \eqref{eq_shape_conc_coupled} is the limit of vanishing viscous friction in the cortex compared to cortex-substrate friction, i.e., $\bar{\eta} \rightarrow 0$.
In this case, the viscous force is negligible and all terms  in the force balance equation \eqref{eqForceBalance} become purely normal which implies the absence of tangential flows.
For $\bar{\eta} \rightarrow 0$ the Jacobian \eqref{eq_Jacobi_full} simplifies to
\begin{align}\label{eq_Jacobi_eta_0}
	\mathbf{J} = \begin{pmatrix}
		 - (l^2 - 1) \bar{\gamma}  & - \bar{V} e^{-\bar{c}_0}  \\
		 (l^2 - 1)  \bar{\gamma} \bar{c}_0 &  \bar{V} e^{- \bar{c}_0} \bar{c}_0 - l^2 \bar{D} - 1
	\end{pmatrix}.
\end{align}
The trace of $\mathbf{J}$ in Eq.~\eqref{eq_Jacobi_eta_0} is given by
\begin{equation}\label{eq_trace_eta_0}
	tr (\mathbf{J}) = \bar{V} e^{- \bar{c}_0} \bar{c}_0 - l^2 \bar{D} - 1 - (l^2 - 1) \bar{\gamma}
\end{equation}
and the determinant by
\begin{equation}\label{eq_determinant_eta_0}
	det (\mathbf{J}) = (l^2 - 1)  ( l^2 \bar{D} + 1 ) \bar{\gamma}.
\end{equation}
Note that the determinant in Eq.~\eqref{eq_determinant_eta_0} does not depend on the polymerization velocity anymore and is always positive. This means that a stationary instability is prohibited.
The base state becomes unstable via a Hopf instability if the trace in Eq.~\eqref{eq_trace_eta_0} changes sign from negative to positive.

These results indicate that actin polymerization is the primary driver of the Hopf instability, while filament diffusion, turnover, and line tension act to stabilize the system. In particular, line tension suppresses deformations of the cell boundary. As $\bar{\gamma}$ increases, Eq.~\eqref{eq_trace_eta_0} shows that a higher polymerization velocity is required to induce the instability. This underscores that cell shape deformations are essential to the emergence of the Hopf instability.
\subsection{Numerical method}
\label{sec:method}
Numerically, the cell cortex is discretized into $N$ nodes at positions $\bm r_i = (x_i, y_i)^T$ ($i = 1,2,...N$) which forms the cell boundary.
We track for each node the actin filament concentration $c_i$ and the velocities which include the cortex velocity $\bm v_{c,i}$, polymerization velocity $\bm v_{p,i}$, and full velocity $\bm v_{i}$ with
$\bm v_{\alpha,i} = (v^x_{\alpha,i}, v^y_{\alpha,i})^T$ ($\alpha \in \{c, p\}$) and $\bm v_{i} = (v^x_{i}, v^y_{i})^T$.
At the beginning of the simulation, the cell shape is set to a circle with radius $R_0$.
The actin concentration is initialized at $c_0$ with an additive perturbation, consisting of the sum over all Fourier modes with random small amplitudes and phases.

The numerical procedure is as follows. 
In each time iteration, we first compute geometry of the cell, see section \ref{sec:numericsGeometry}, which serves as input for all other steps.
We then compute the nodal forces using weak formulation and solve the force balance to obtain the cortex velocity at each node, see section \ref{sec:numericsForces}.
The now obtained full velocity allows us to compute the concentration evolution at each node in section \ref{sec:numericsAdvDiff}.
We use an explicit Euler scheme to evolve the node positions according to the full velocity and concentrations in time with time step $\Delta t$.
In a final step, outlined in section \ref{sec:numericsMeshAdvection}, we tangentially shift mesh nodes to avoid distortions of the mesh and update the actin concentration accordingly.
\subsubsection{Geometry}
\label{sec:numericsGeometry}
In each time step, we first compute key geometric quantities.
The edge length is given by $l_i = \abs{\bm r_{i} - \bm r_{i-1}}$.
The global arc length at node $i$ is given by
\begin{equation}
	s_i = \sum_{j = 1}^i l_i.
\end{equation}
The tangential unit vector at node $i$ is computed as
\begin{equation}
	\bm \tau_i = \frac{\bm r_{i+1} - \bm r_{i-1}}{\abs{\bm r_{i+1} - \bm r_{i-1}}}.
\end{equation}
The unit normal vector $\bm n_i$ is obtained by rotation of $\bm \tau_i$ around $- \frac{\pi}{2}$.
We further compute the cell area and the cell perimeter.
\subsubsection{Force balance and cortex velocity}
\label{sec:numericsForces}
Continuing, we compute the forces acting on the cortex using the weak formulation.
The nodal basis function is given by
\begin{align}\label{eqDefHatFunction}
	\varphi_i (s) = \begin{cases}
		\frac{s - s_{i-1}}{l_i} & \text{for} \quad s \in [s_{i-1}, s_i] \\
		\frac{s_{i+1} - s}{l_{i+1}} & \text{for} \quad  s \in [s_{i}, s_{i+1}] \\
		0 & \text{otherwise} 
	\end{cases}
\end{align}
With this and by using a trapezoidal quadrature rule, the tension force at node $i$ is obtained as
\begin{equation}\label{eq:tensionForceWeak}
	\bm f^\text{tens}_i = - 2 \gamma \frac{p_\text{max} - p_r}{p_\text{max} - p} \frac{ \left( \frac{\bm r_{i} - \bm r_{i-1}}{l_i} + \frac{\bm r_{i} - \bm r_{i+1}}{l_{i+1}}  \right) }{ l_{i} + l_{i+1} }.
\end{equation}
Similarly, it follows for the viscous force at the $i$-th node
\begin{align}\label{eqViscousForceWeakFinal}
	\bm f^\text{visc}_i = \notag - \frac{2 \eta}{l_i + l_{i+1}} &\left[ \frac{(v^x_{c,i} - v^x_{c,i-1})(x_i - x_{i-1}) + (v^y_{c,i} - v^y_{c,i-1})(y_i - y_{i-1})}{l_i^3} \begin{pmatrix}
		x_i - x_{i-1} \\ y_i - y_{i-1}
	\end{pmatrix} \right. \\
	&\left.- \frac{(v^x_{c,i+1} - v^x_{c,i})(x_{i+1} - x_{i}) + (v^y_{c,i+1} - v^y_{c,i})(y_{i+1} - y_{i})}{l_{i+1}^3} \begin{pmatrix}
		x_{i+1} - x_{i} \\ y_{i+1} - y_{i}
	\end{pmatrix}  \right].
\end{align}

Now that all forces are obtained, we solve the force balance equation for the cortex velocity.
Since both the viscous and the drag force depend on the cortex velocity, the force balance is an implicit equation.
Rather than solving it directly, we solve
\begin{equation}\label{eqForceBalanceMassWeak}
	-\zeta \bm v_{c, i} + \bm f_i^\text{visc} + \bm f_i^\text{tens} + \bm f_i^\text{area} = m_i \frac{d \bm v_{c,i}}{dt}
\end{equation}
for the cortex velocity.
The right-hand side of Eq.~\eqref{eqForceBalanceMassWeak} represents an inertial force line density, where $m_i$ is a small but finite mass line density.
From this results the update rule for the  cortex velocity,
\begin{equation}\label{eqUpdateCortexVelocity}
	\bm v_{c,i} (t + \Delta t) = \bm v_{c,i} (t) + \left[ -\zeta \bm v_{c, i} (t) + \bm f_i^\text{visc} (t) + \bm f_i^\text{tens} (t) + \bm f_i^\text{area} (t) \right] \frac{\Delta t \left[l_i(t) + l_{i+1}(t)\right]}{2 M}
\end{equation}
where 
\begin{equation}\label{eqInertMass}
	M = \int_{\Gamma} m \ d s \approx \sum_{i = 1}^{N} m_i \frac{l_i + l_{i+1}}{2}.
\end{equation}
is the total inert mass along the contour.
In our simulations we choose as parameters either
$\Delta t = 10^{-6}$, $m =  10^{-5}$, and $N = 100$ or
$\Delta t = 10^{-7}$, $m =  10^{-6}$, and $N = 200$, 
depending on the required spatial resolution of the cortex.
\subsubsection{Advection-diffusion equation}
\label{sec:numericsAdvDiff}
Numerically, we track filament labels along the cortex.
We thus solve the advection diffusion equation in the filament co-moving frame,
\begin{align}\label{eqAdvDiffusionWeak}
	\int_{\Gamma} \frac{D c}{D t} \ \varphi_i dr = \int_{\Gamma} \left[ - (\bm \nabla_\Gamma \cdot \bm v) c + D \Delta_\Gamma c + \beta (c_0 - c) \right] \varphi_i ds.
\end{align}
A straightforward calculation yields
\begin{align}\label{eqDiffusionTermWeak3}
	\notag
	\frac{D c}{D t}_i = &- c_i \frac{\frac{(v^x_{i} - v^x_{i-1})(x_i - x_{i-1}) + (v^y_{i} - v^y_{i-1})(y_i - y_{i-1})}{l_i} + \frac{(v^x_{i+1} - v^x_{i})(x_{i+1} - x_{i}) + (v^y_{i+1} - v^y_{i})(y_{i+1} - y_{i})}{l_{i+1}}}{l_i + l_{i+1}}\\ 
	&- 2 D  \frac{ \frac{c_i - c_{i-1}}{l_i} - \frac{c_{i+1} - c_{i}}{l_{i+1}}}{l_i + l_{i+1}} + \beta (c_0 - c_i),
\end{align}
where
\begin{equation}\label{eq:fullVelNode}
	\bm v_i = \bm v_{c,i} + V e^{- \frac{c_i}{c_r}}
\end{equation}
is the full velocity at the $i$-th node.
\subsubsection{Mesh advection}
\label{sec:numericsMeshAdvection}
Updating node positions  using the full velocity $\bm v_i$ leads to strong inhomogeneity in the node distribution along the cortex which decreases numerical accuracy.
To prevent this, we apply a tangential advection of the mesh nodes ensuring their approximately uniform distribution along the contour.
This is implemented by evolving the node positions according to
\begin{align}\label{eqUpdateNodePosAdvection}
	\bm r_i (t + \Delta t) = \bm r_i (t) + \Delta \bm r_i^\text{adv} (t).
\end{align}
Herein,
\begin{align}\label{eqMeshAdvectionField}
	\Delta \bm r_i^\text{adv} = \chi \left[ \left( \bm r_{i+1} - 2 \bm r_{i} + \bm r_{i-1} \right)  \cdot \bm \tau_i \right] \bm \tau_i
\end{align}
is a tangential increment which tends to minimize the distances of node $i$ to its neighbors where $\chi$ is a stiffness parameter.

The tangential shift of mesh nodes requires the interpolation of the actin concentration accordingly.
In linear approximation the concentration increment is given by
\begin{equation}\label{eq:concentrationAdvection}
	\Delta c = \Delta \bm r_i^\text{adv} \cdot \bm \nabla_\Gamma c.
\end{equation}
We found that the direct update given in Eq.~\eqref{eq:concentrationAdvection} can lead to negative values for the concentration.
We prevent this by updating an auxiliary variable defined for each node as $\psi_i = \ln (c_i)$.
The concentration at the $i$-th node that corresponds to its position upon tangential advection is then given by
\begin{equation}
	c_i^\text{advected} = e^{\psi_i + \Delta \psi_i}
\end{equation}
with the increment of the auxiliary variable
\begin{align}\label{eqWeakFormMeshAdvectionFinalPsi}
	\Delta \psi_i = \frac{(\psi_{i} - \psi_{i-1}) \left( \frac{\bm r_i - \bm r_{i-1}}{l_i} \right) \cdot \Delta \bm r^\text{adv}_i + (\psi_{i+1} - \psi_{i}) \left( \frac{\bm r_{i+1} - \bm r_{i}}{l_{i+1}} \right) \cdot \Delta \bm r^\text{adv}_i}{l_i + l_{i+1}}.
\end{align}
In our simulations we choose $\chi = 0.25$.
\subsection{Influence of actin diffusivity on circular migration}
\label{sec:circularDiffusivity}
\begin{figure}[h]
	\includegraphics[width=0.5\columnwidth]{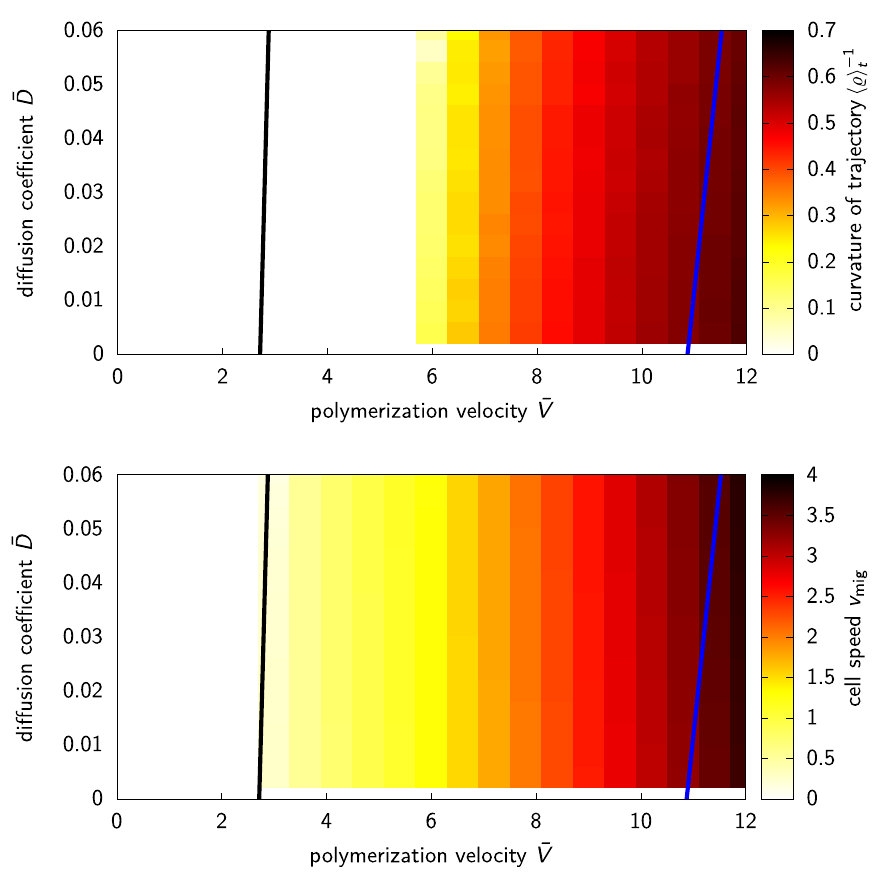}
	\caption{Phase diagrams showing (top) the trajectory curvature $\langle \varrho \rangle_t^{-1}$ and (bottom) the cell speed (color code, see legend) as a function of the dimensionless diffusion coefficient $\bar{D}$ and polymerization velocity $\bar{V}$.
		Black (blue) line shows the analytically derived threshold of the first (second) harmonic.
	}
	\label{fig:circularPhaseDiagDiffCoeff}
\end{figure}

We study the effect of cortical filament diffusivity on cell migration. 
Fig.~\ref{fig:circularPhaseDiagDiffCoeff} (top) shows the curvature of cell trajectories as a function of the diffusion coefficient and the polymerization velocity.
We find that $\langle \varrho \rangle_t^{-1}$ decreases for increasing $\bar{D}$.
This means that filament diffusion acts against the tendency for circular motion.
Fig.~\ref{fig:circularPhaseDiagDiffCoeff} (bottom) shows that the migration speed increases as a function of $\bar{V}$ for all values of $\bar{D}$.

\end{document}